\documentclass[sigconf, noacm]{aamas} 
\subtitle{A Preprint}

\renewcommand\footnotetextcopyrightpermission[1]{} 
\makeatletter
\renewcommand\@formatdoi[1]{\ignorespaces}
\makeatother

\usepackage{balance} 
\usepackage{booktabs} 
\usepackage{graphicx}
\usepackage{microtype}
\usepackage{caption}
\usepackage{subcaption}
\usepackage[flushleft]{threeparttable}
\usepackage{multirow}
\usepackage{balance}
\usepackage[inline]{enumitem}

\usepackage{float} 
\usepackage{hyperref} 
\usepackage{siunitx}
\usepackage{array}
\usepackage{verbatim}
\DeclareMathOperator*{\argmax}{arg\,max}
\DeclareGraphicsExtensions{.pdf,.jpeg,.png,.jpg}

\usepackage{xspace}
\newcommand{\eg}{e.g.,\xspace}
\newcommand{\etal}{et al.\xspace}
\newcommand{\ie}{i.e.,\xspace}
\newcommand{\etc}{etc.\xspace}

\newcommand{\fig}{Figure\xspace}
\newcommand{\sect}{Section \xspace}

\setcopyright{none}
\acmConference[AAMAS '21]{Proc.\@ of the 20th International Conference on Autonomous Agents and Multiagent Systems (AAMAS 2021)}{May 3--7, 2021}{Online}{U.~Endriss, A.~Now\'{e}, F.~Dignum, A.~Lomuscio (eds.)}
\copyrightyear{2021}
\acmYear{2021}
\acmDOI{}
\acmPrice{}
\acmISBN{}

\title[CHARET: Character-centered Approach to Emotion Tracking in Stories]{CHARET: Character-centered Approach\\ to Emotion Tracking in Stories}

\author{Diogo S. Carvalho}
\affiliation{
  \institution{INESC-ID \& Instituto Superior Técnico, Universidade de Lisboa}
  }
\email{diogo.s.carvalho@tecnico.ulisboa.pt}

\author{Joana Campos}
\affiliation{
  \institution{INESC-ID}
  }
\email{joana.campos@inesc-id.pt}

\author{Manuel Guimarães}
\affiliation{
  \institution{INESC-ID \& Instituto Superior Técnico, Universidade de Lisboa}
  }
\email{manuel.m.guimaraes@tecnico.ulisboa.pt}

\author{Ana Antunes}
\affiliation{
  \institution{INESC-ID \& Instituto Superior Técnico, Universidade de Lisboa}
  }
\email{ana.j.antunes@tecnico.ulisboa.pt}

\author{João Dias}
\affiliation{
  \institution{INESC-ID \& Faculdade de Ciências e Tecnologia, Universidade do Algarve \& CCMAR}
  }
\email{jmdias@ualg.pt}

\author{Pedro A. Santos}
\affiliation{
  \institution{INESC-ID \& Instituto Superior Técnico, Universidade de Lisboa}
  }
\email{pedro.santos@tecnico.ulisboa.pt}

\begin{abstract}
Autonomous agents that can engage in social interactions with a human is the ultimate goal of a myriad of applications. A key challenge in the design of these applications is to define the social behavior of the agent, which requires extensive content creation. In this research, we explore how we can leverage current state-of-the-art tools to make inferences about the emotional state of a character in a story as events unfold, in a coherent way. We propose a character role-labelling approach to emotion tracking that accounts for the semantics of emotions. 
We show that by identifying actors and objects of events and considering the emotional state of the characters, we can achieve better performance in this task, when compared to end-to-end approaches.
\end{abstract}

\newcommand{\BibTeX}{\rm B\kern-.05em{\sc i\kern-.025em b}\kern-.08em\TeX}

\begin{document}


\pagestyle{fancy}
\fancyhead{}


\maketitle
\thispagestyle{empty}

\section{Introduction} 
Intelligent Virtual Agents (IVA) have an ever increasing range of applications from conversational interfaces on websites to tutors or teammates in educational environments \cite{pecune2016evaluating, christoffersen2002make}, where they are equipped with tools to conduct human-like interactions, in closed-context environments. It is in the role of commercial automated assistants (\eg Siri, Alexa, Google Home) that IVAs are currently the most popular. Their conversational skills are a result of technological advances in Natural Language Processing (NLP) that allow IVAs to support the user in everyday tasks. Although these systems are getting more and more sophisticated, their communication abilities are still limited. These end-to-end deep learning approaches to dialog generation for IVAs focus only on response quality and do not explicitly control social factors in natural human-like interactions. Among other aspects, to be \textit{socially resonant} \cite{kopp2010social}, IVAs have to understand the user's beliefs, emotions, goals and intentions, while maintaining and sharing their own, to produce consistent behaviour overtime. 

 Building computer artifacts that engage in the complex social dance, particularly in open-ended domains, is a compelling prospect that has attracted many researchers. Agent-based attempts to simulate individual cognitive and affective processes \cite{dias2005feeling, hartholt_all_2013, popescu2014gamygdala, marsella2009ema}, model how individual traits, goals, beliefs and actions interact to produce intelligent and emotionally plausible behaviour, in any scenario that the user can imagine. Yet, it is up to the author of a scenario to manually describe them for each character and guarantee plot adaptability and consistency as events unfold. While this can be manageable in narrow domains of application, scenario complexity can escalate rapidly, reducing the power of such architectures by relying heavily on the accessibility of the tool \cite{mascarenhas2018virtual} and foremost the authors' creativity and ability to anticipate all interaction paths. 
 
In our project, 
the ultimate goal is to create agents that can effectively act socially on the users' actions, without relying on hand-generated content only. At the same time. it is our stance that we cannot depend solely on machine learning to create socially resonant agents and we need to leverage symbolic models, semantic networks and other conceptual models to create more natural interactions. In this work, we explore this idea by focusing on a central element of social interactions -- \textit{emotions}. We are interested in detecting and understanding the user's (or a character) emotional state, as interactions evolve in ways that were not accounted for. 
To that end, we explore how we can leverage state-of-the-art tools to accurately infer emotions of characters in stories, as events unfold. 

In this paper, the emotion classification task is modeled as a character role-labeling problem, because we are interested in who felt the emotion and why. Our character-centered approach diverges from common approaches to sentiment analysis in text that attempt to infer an emotion from a set of words, ignoring the semantics and subjectiveness of emotions: \textit{emotional reactions are caused by an event that the actor and the object (of that event, if exists) may experience differently}. Furthermore, in our perspective emotions are not static reactions to events, but \textit{dynamic constructs that evolve as interactions unfold}. We find that by identifying the characters' roles \-- actors and objects of an event \-- in a story and keeping track of their emotional state, we can perform better than end-to-end approaches in the task of emotion tracking in stories. We draw from the results of this work particular challenges for domain authoring of open-ended socially resonant human-agent interactions.
 
%
\section{Related Work}
Research in psychology, neurology and cognitive science shows that not only do people use their cognitive functions, but they also heavily rely on their emotions when taking decisions \cite{puica2013emotional}. Damasio \etal proved that if these two parts don’t interconnect in a proper manner, multiple options are harder to be filtered and bad decisions are easier to take \cite{marg1995descartes}. These findings influenced the design of intelligent characters and led agent-based modelling systems to build their frameworks around emotions. Affective Agents not only aim at being more realistic and providing a more engaging experience in human-computer interaction, but also at improving the performance of rational agents. 


Other works have explored the extraction of sentiment from characters in stories. Among others discussed in Section \ref{sec:baselines}, Paul \etal explored how to track the needs of such characters by leveraging commonsense embedded in knowledge graphs \cite{paul2019ranking} and also how to incorporate attention models and semantic role-labeling to perform emotion classification tasks \cite{paul2020social}. Gaonkar \etal aimed at capturing the semantics of emotion labels and showed the benefits of such approach over the standard \cite{gaonkar2020modeling}. In Section \ref{subsec:expeval-approach} we introduce and discuss our proposed approach.

\subsection{Affective Agent's Modelling Architectures}

FAtiMA Toolkit is a collection of open-source tools that is designed to facilitate the creation and use of cognitive agents with socioemotional skills \cite{guimaraes2019accessible}. Its objective is to help researchers, developers and roboticists to incorporate a computational model of emotion and decision-making in their projects. In particular, it enables developers to easily create Role Play Characters. These are socially intelligent characters with detailed AI modules that makes them autonomous regarding social interactions \cite{westera2020artificial}. Both the Emotional and the Decision Making processes behind FAtiMA-designed characters are defined by logical rules. Expressing the agent's decisions using conditions with logical variables the action space of the agent grows and adapts according to its beliefs \cite{mascarenhas2018virtual}.

The Virtual Human Toolkit \cite{hartholt_all_2013} is a well known architecture that is also designed to facilitate the creation of autonomous conversational characters. Moreover, the architecture is also highly modular. In this case, the modules that are provided are focused in handling aspects that are more related to the embodiment of a character rather than its cognitive abilities. 
This functionality includes aspects and services such as
speech processing, emotional modeling of the learner, emotional modeling of the virtual human, the gestures of the virtual human, rendering, and other services \cite{brawner2016agent}.

GAMYGDALA \cite{popescu2014gamygdala} is a computational model of emotions that is based on the OCC theory\cite{clore2013psychological}. Similar to FAtiMA Toolkit, this emotional appraisal engine was designed to be more accessible to game developers. Essentially, authors need only to provide a list of goals for each character and then specify which events will block or facilitate each goal. Based on that information, the engine will determine the changes made to the character's emotional state \cite{endo2019developing}.


As we have mentioned before,  agent modelling architectures rely on a type of authoring that is oriented towards cognitive concepts such as goals, beliefs and emotions, among others \cite{mascarenhas2018virtual}. In turn, they are also heavily dependent on the designer’s ability to imagine a variety of social situations and to use their intuition to specify social behaviours and execution rules for the agent, which may be difficult to articulate \cite{liu2016data}, particularly for people outside of the scientific field. 

In the past, when authoring Agent Modelling Architectures, this task would be manually performed by a team of engineers or by the developers of the architecture themselves.  However, recent developments in Natural Language Processing (NLP) and Machine Learning (ML) fields have led to the automatization, or at least partial automatization, of this task.

One of the most recent and successful approaches to this problem is the translation of natural language descriptions into agent-readable concepts. Authors are asked to provide textual input such as stories or scrips and the system is able to compute and transform it into intelligent character's scenarios.


For example, in the AI Planning Field, much like in the IVA field, there is the underlying assumption that users can formulate the problem using some formal language \cite{lindsay2017framer}. Here, knowledge acquisition tools have been used to extract the domain model through NLP. Framer uses Natural Language descriptions, written by users, as input and is able to learn planning domain models \cite{feng2018extracting}. Janghorbani et al. \cite{janghorbani2019domain} introduced an authoring assistant tool to automate the process of domain generation from natural language description of virtual characters. 

Given the importance of emotions on intelligent agent modelling systems, we believe that the first step towards easing the authoring burden should focus in detection of emotion in Natural Language texts describing a story or sequence of events. This will allow us to replace traditional hand-authored appraisal rules used in socio-emotional agent systems \cite{guimaraes2019accessible} by a learned model that is able to subjectively appraise an event according to different perspectives.

\subsection{Emotion Classification of text}
Sentiment analysis is the umbrella term for a series of tasks focused on detecting valence, emotions and other affective states in text. Some tasks explore the positive or negative orientation of words \cite{wilson2005recognizing}, while other intend to infer a driving sentiment or opinion in a whole document \cite{bakshi2016opinion}. These works focus on extracting an emotional label (or overall sentiment) from a set of words and do not consider the emotional state of the different entities in the text. Different methods including supervised machine learning, lexicon-based approaches and linguistic analysis are the common techniques used to tackle sentiment analysis of text, but recently developed tools have open avenues for sentiment detection in text. In particular, a newly created commonsense tool \--ATOMIC \cite{sap2019atomic} \-- that allows to reason about causes and effects of events has shown its applicability in question answering (QA) tasks \cite{sap2019socialiqa} and the potential to reason about events and their related emotions. While emotion classification of events can be considered a QA task under this setting, it is important to note that an individual's emotional reaction to an event is not static and may depend on other aspects. As shown by \citeauthor{rashkin2018modeling} \cite{rashkin2018modeling} story context is relevant for emotion tracking in stories and methods should be devised to capture it. More recently \cite{bosselut2019dynamic}, the power of Transformer language model tools allowed the creation of COMET \cite{bosselut2019comet}, a new framework for training neural representations of knowledge graphs that leverages ATOMIC knowledge base. This new tool allows to make contextualized inferences (using the information in the graph), which are the scaffold of a character's emotional state classification.

Given the underlying semantics of emotions, we consider that an important form of context is \textit{who} felt what given a certain event. As noted by others \cite{mohammad2014semantic} identifying semantic roles in a sentence can improve sentiment analysis task because sentiment is not always explicit in text. This implies that some pre-processing is necessary and that end-to-end approaches may not suffice in this task \cite{cambria2017sentiment}.


%

\section{Research Question} \label{sec:research_q}
Following the related work we are introducing a new mechanism that considers the semantics of emotions to track and predict the emotional state of a character in a story. We consider that a character-centered approach to emotion classification
that leverages state-of-the-art commonsense tools designed around cognitive processes ( \eg beliefs, emotions, intentions, causes and effects) will assist in the task of creating of intelligent character's scenarios. This consideration drives the research questions for the evaluation latter presented in the paper:

\begin{itemize}
 \item[\textbf{RQ1}] \textit{Can state-of-the-art commonsense tools allow an agent to keep track of the emotional state of a character as events unfold, in a coherent way?}
\end{itemize}

This question can be broken down into more specific sub-questions:
\begin{enumerate}
    \item Is it possible to make a better use of commonsense inference tools by considering the semantics of emotions?
    \item Does a layered approach to emotion tracking, \ie an approach that breaks down the problem into meaningful sub-problems, yield better results than end-to-end approaches?
\end{enumerate}

This research question seeks to address whether character specific inferences and context improve emotion classification in stories, when compared to previous work  \cite{bosselut2019dynamic} that use the same context across characters, when making new inferences. A layered approach to sentiment analysis refers to the set of pre-processing steps required to apply the proposed approach. These steps are detailed in \sect \ref{subsec:expeval-approach} and constitute the approach pipeline.




\section{CHARET: Character-centered Approach to Emotion Tracking in Stories}\label{subsec:expeval-approach}
In this paper we propose a \textbf{character-centered} approach to emotion recognition and tracking in stories that identifies stimuli and their objects, aided by semantic role-labelling \cite{jurafsky2009language}. We formalize it in the following way. Given a story $S$, consisting of $n$ events, $(s_1, \ldots, s_n)$, and a set of $m$ characters $C = \{c_1, \dots, c_m\}$, we assume that emotional episodes are defined by an event-character pair $(s_t, c_i)$ in $S \times C$, which have a corresponding set of emotional reactions  $Y_{s_t,c_i}$. The set $Y_{s_t,c_i}$ is a subset of $P = \{emotion_1, \dots, emotion_N\}$ consisting of a previously defined set of emotions. We highlight that the empty set, corresponding to \textit{no emotion}, and the whole set, corresponding to \textit{all emotions}, are allowed. Here, we follow OCC theory of emotions\cite{ortony1998}, which posits that multiple emotions can be experienced simultaneously as the result of an appraisal of an event. Possible choices of $P$ include Ekman's basic emotions \cite{ekman1992there} or Plutchik's wheel of emotions \cite{plutchik1980general}.

Provided S and C, we set ourselves the task of tracking the emotions in $Y_{s_t,c_i}$ with the support of commonsense inference tools. Our approach is as follows. For each event-character pair $(s_t, c_i)$, for each emotion $y$ in $P$, a function $f:S \times C \times P \to \{0, 1\}$ predicts whether $y$ is an emotional reaction of the state-character pair $(s_t, c_i)$, or in other words whether $y \in Y_{s_t, c_i}$, based on a score, $score_{s_t,c_i,y}\in [0, 1]$. If the score is sufficiently high, we classify $y$ as an emotional reaction to $(s_t, c_i)$. Once we followed the approach for every emotion $y$ from $P$, we are left with the predicted set of emotional reactions $\hat{Y}_{s_t, c_i}$.

The function $f$ encompasses a pipeline with three components:
\begin {enumerate*} [label=\itshape\alph*\upshape)]
\item character role-labeling, 
\item commonsense inference, and 
\item emotion classification.
\end {enumerate*} It consists in 
identifying stimuli (events) and their objects, and then use a language model \--- \textit{COMET} \cite{bosselut2019comet} \--- to make inferences about life events and identify the triggered emotions, depending on the perspective of the character in a story. We detail our pipeline in the following subsections. 

\begin{figure}[h]
  \centering
  \includegraphics[width=\linewidth]{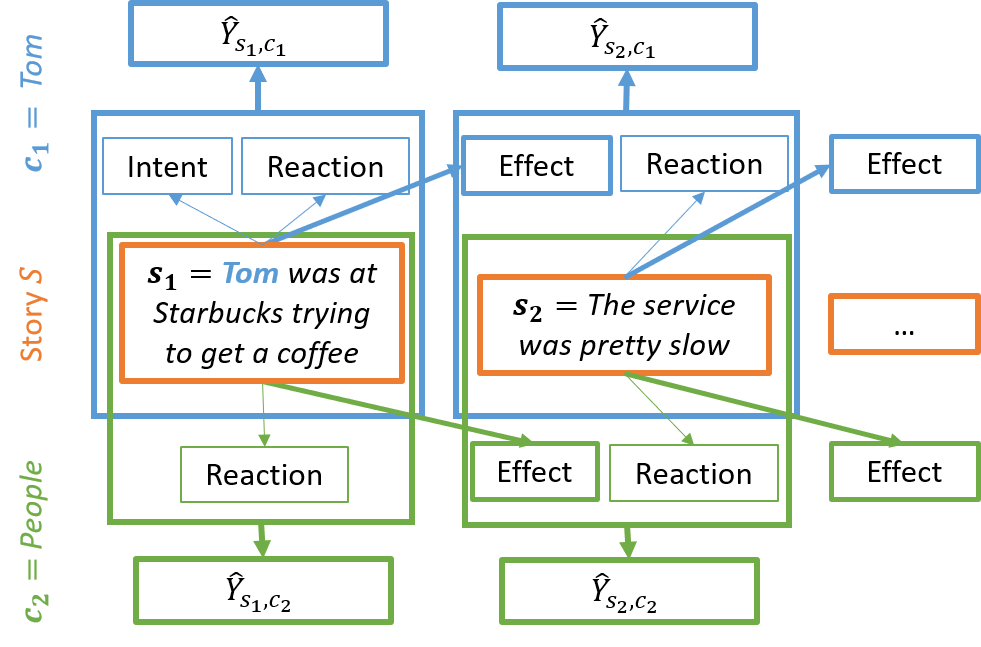}
  \caption{Diagram of our approach. the inferences of \textit{effects} from the first event are used as context to classify the emotions of the characters on the second event.}
  \label{fig:diagram}
  \Description{}
\end{figure}

To capture \textbf{character-specific context} and allow to detect more coherent emotions as events unfold, we use information from the previous event $s_{t-1}$. Specifically we consider the \textit{effects} that the previous event, $s_{t-1}$, had on each character, to model how the character feels about $s_t$. This entails that emotions are not momentary reactions to an event, but instead are constructs that incrementally unfold. Figure \ref{fig:diagram} shows a diagram of our approach over the first two examples of a story from StoryCommonsense\footnote{\url{https://uwnlp.github.io/storycommonsense/}}, \textit{Hot Coffee}.




\subsection{Character Role-Labeling}

To establish a character's role with respect to an event in a sentence \--- a character can be either the \textit{actor} or the \textit{object} of an event \--- we use PredPatt. PredPatt \cite{white-EtAl:2016:EMNLP2016, zhang-EtAl:2017:IWCS} is a tool that can be used to perform semantic-role labeling, since it defines a set of interpretable, extensible and non-lexicalized patterns based on Universal Dependencies \cite{de2014universal, droganova2019towards} and extracts predicates and their arguments using these rule-based patterns. While other (deep) tools, such as AllenNLP semantic role-labelling \cite{gardner2018allennlp}, are available to perform this task, it has been shown in the literature that rule-based system produce better results \cite{xiang2019survey}. PredPatt is an attractive tool to our project because it allows to perform SRL in other languages, different than English, without consequences to the other blocks in the pipeline.

\begin{figure}[h]
  \centering
  \includegraphics[width=\linewidth]{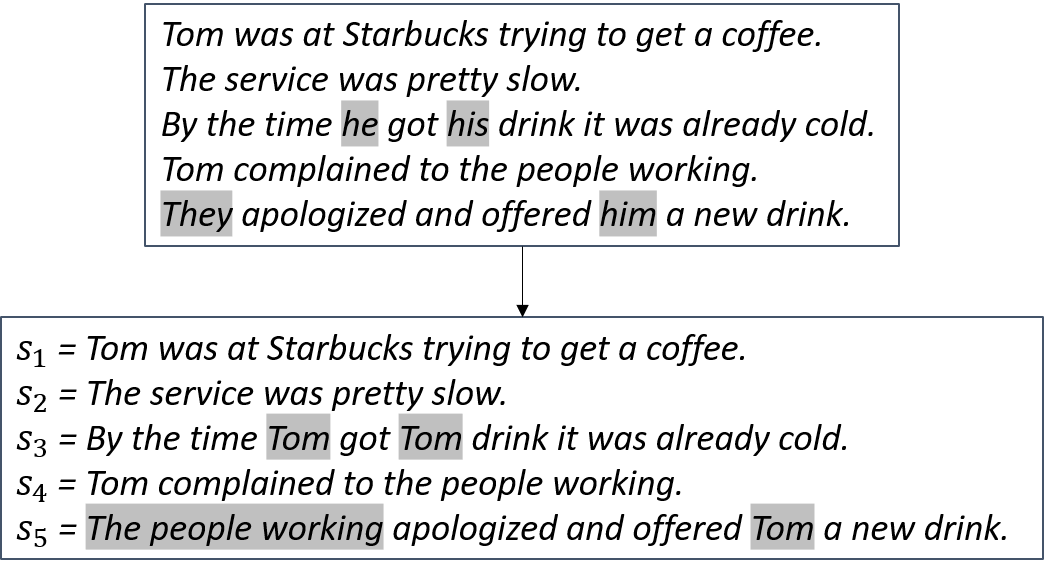}
  \caption{Co-reference solution example. The box on top is a five-line story present in the StoryCommonsense dataset. The box at the bottom demonstrates how the co-references were resolved.}
  \label{fig:coref}
  \Description{}
\end{figure}

\begin{figure}[h]
  \centering
  \includegraphics[width=\linewidth]{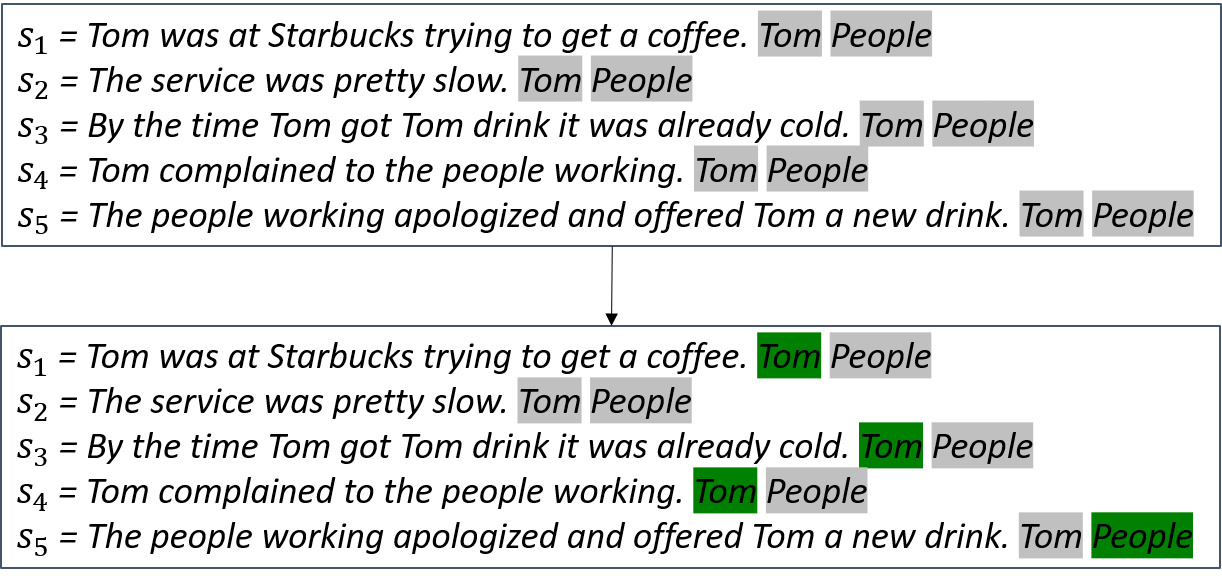}
  \caption{Character Role-Labeling example. Given the two entities present in this story \--- \textit{Tom} and \textit{People} -- the algorithm identifies who is the \textit{actor} of the event (in green).}
  \label{fig:rolelab}
  \Description{}
\end{figure}

Note that as a pre-processing step we had to resolve co-references within a story (see \fig \ref{fig:coref}). We used NeuralCoref\footnote{\url{https://github.com/huggingface/neuralcoref}} to assist in this task. NeuralCoref annotates and resolves co-reference clusters using a neural network.  The system receives a set of ordered sentences, which constitute a story, and substitutes the pronouns \textit{he}, \textit{his}, \textit{they} and \textit{him} by the corresponding entity  (see \fig \ref{fig:rolelab} ). From there, we use PredPatt to classify the character with respect to its role on the event and keep track of the characters in the story.
%


%

%


%

\subsection{Commonsense Inference}\label{sec:commonsense}
COMET is a tool for automatic knowledge graph construction. It is constructed as a Transformer language model fine-tuned on the ATOMIC knowledge graph. In ATOMIC, each event is annotated with possible intents (\textit{xIntent}), needs (\textit{xNeed}), reactions (\textit{xReact}), attributes (\textit{xAttr}) and effects (\textit{xWant, xEffect}), with respect to the \textit{actors} of the event. With respect to others, each event is annotated with possible reactions (\textit{oReact}) and effects (\textit{oWant}, \textit{oEffect}).
Given an event, COMET is thus able to perform commonsense inferences about events described in natural language. Such inferences can, again, include intents, attributes, effects and, even more importantly for our work, emotional reactions. 

\begin{figure}[h]
  \centering
  \includegraphics[width=4.5cm]{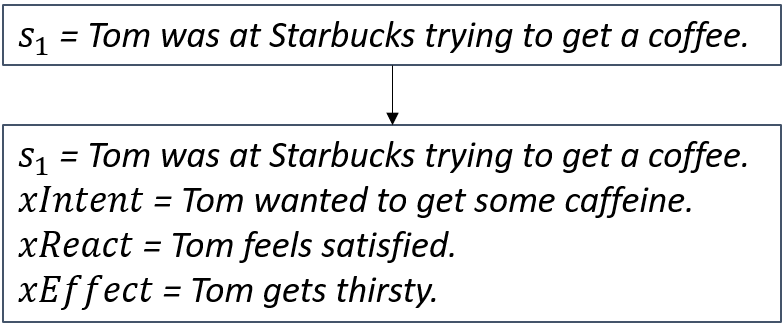}
  \caption{Commonsense Inference example. Given $s_1$ we infer the most likely intent, reaction and effect of the character \textit{Tom}, using COMET.}
  \label{fig:commoninf}
  \Description{}
\end{figure}

We build on the work of \citeauthor{bosselut2019dynamic} \cite{bosselut2019dynamic}, who leveraged commonsense inference to reason about unstated events in question answering tasks. In the cited work, the author's propose COMET-CGA, an approach which uses COMET to reason over commonsense inferences.
For each story event-character pair $(s_t, c_i)$, we perform commonsense inferences to produce unstated events, using COMET \cite{bosselut2019comet}.
In our approach, if a character is an actor of the event, we use COMET to infer its most likely intent, reaction and effect, producing a set of character-specific events \[E_{s_t,c_i}^x = \{s_t, xIntent, xReact, xEffect\}.\] If, on the contrary, a character is an object of the event, we use COMET to infer only its reaction and the effect, producing the set of character-specific events \[E_{s_t,c_i}^o = \{s_t, oReact, oEffect\}.\] \fig \ref{fig:commoninf} shows an example of the commonsense inference step of our pipeline.

As stated before, to capture character-specific context we use the effect inferences -- both \textit{xEffect} and \textit{oEffect} -- from the previous story line, $s_{t-1}$. If $t = 0$, we do not use any \textit{effect} inference. The set of character-specific events $E_{s_t,c_i}$ can be seen as the set of character-specific information available. Each piece of information is described as an event described in natural language. While the attempt to add story context to emotion classification in stories is not novel (see \sect \ref{sec:baselines}), using unstated inferred events from previous events is. Adding story context intends to classify emotions more coherently throughout the story. 

\subsection{Emotion Classification}
While built with the aim of commonsense inference for knowledge graph construction, COMET can also be used in emotion classification tasks (see \sect \ref{sec:baselines}). The use of COMET for emotion classification is not new, and the main difference in our approach is that, by including a Character Role-Labeling step, we can make character-specific commonsense inferences from events and classify different emotions for different characters, according to their role.

Being a language model, once an event is inputted, to produce a reaction inference, COMET uses a probability distribution over a vocabulary V, corresponding to the probabilities of outputting each word in V. As such, and for example, we can model the likelihood that \textit{surprise} is an emotion of a certain event-character pair as the probability $p_{e, surprise}$ of the first word of the reaction inference being \textit{surprised}, or the likelihood of \textit{joy} as $p_{e, happy}$. More generally, we can define a dictionary $d$ that, for each emotion $y$ in $P$, maps the emotion to a word from the vocabulary $V$. The choice of dictionary is ad-hoc and can there are various alternatives for each set $P$ of emotions considered.

For each event $e$ in the set $E_{s_t,c_i}$, and for each emotion $y$ in $P$, we compute the probability $p_{e, y}$ that emotion $y$ is the reaction inference from event $e$. We observe that, if for the event-character pair $(s_t, c_i)$, the character $c_i$ is classified as an actor in the Character Role Labeling step, we use the probabilities from COMET's \textit{xReact} inference; if, on the other hand, the character $c_i$ is an object, we use the probabilities from COMET's \textit{oReact} inference. Finally the score of the emotion is the geometric mean of these probabilities, \[score_{s_t,c_i,y} = \sqrt[\leftroot{-2}\uproot{2}|E_{s_t,c}|]{\prod_{e\in E_{s_t,c_i}} p_{e,y}},\] same as in COMET-CGA. Other alternatives to the geometric mean exist, such as the arithmetic mean, the maximum and the minimum.
Based on the emotion score, we must decide whether the emotion $y$ is a reaction of character $c_i$ and event $s_t$. We establish that $y$ is an emotional reaction of the event-character pair $(s_t, c_i)$ if $score_{s_t, c_i, y}$ is bigger than an emotion specific threshold $k_y\in[0, 1]$. Conversely, we establish that $y$ is not an emotional reaction of the event-character pair if $score_{s_t, c_i, y}$ is lower or equal than the emotion specific threshold $k_y$. The emotion specific thresholds $k_y$ allow to make up for any bias towards more frequent emotions such as \textit{happy} and \textit{sad}, as opposed to \textit{fearful} and \textit{trusting}, since we can use lower thresholds for the least frequent emotions.

\begin{figure}[h]
  \centering
  \includegraphics[width=\linewidth]{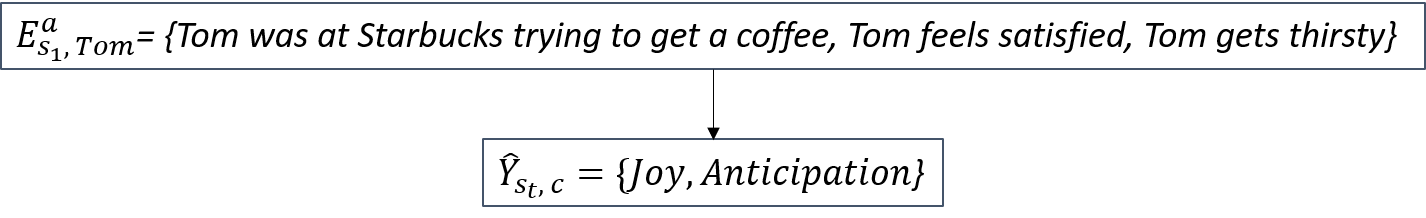}
  \caption{Emotion Classification example. Each event on the set on top gives each emotion a probability. For each emotion, the probabilites from each event are combined to produce a score. The emotions with a score bigger than its specific threshold are classified. In this case, the classified emotions are \textit{Joy} and \textit{Anticipation}.}
  \label{fig:emoclass}
  \Description{}
\end{figure}

A character's emotion classification can be executed under two settings: zero-shot and one-shot, as can COMET-CGA  \cite{bosselut2019dynamic}. In the zero-shot setting, the emotion specific thresholds are not optimized, to allow for generality of the approach and ease applicability. In the one-shot setting, each emotion specific threshold is optimized by sweeping a previously defined space of possible thresholds. In the zero-shot setting, we define each emotion threshold as a percentile. Specifically, if an emotion $y$ appears on $q\%$ of the training set, we define $k_y$ as the $q$-th percentile of the cumulative distribution function of scores for emotion $y$. In the few-shot setting, we define each emotion threshold as the lowest value maximizing the F1-score on the training set. 
Figure \ref{fig:emoclass} shows an example of the emotion classification step.



\section{Experimental evaluation}\label{sec:expeval}
In this section we describe the conducted experimental evaluation in StoryCommonsense dataset \cite{rashkin2018modeling}, which was annotated using a character-centered approach. The semantics of emotions were taken into account in the annotation of this dataset, allowing access to emotional states from the point-of-view of the actor and object of an event. We test and compare CHARET against previous approaches, which are discussed below (see \sect \ref{sec:baselines}).


\subsection{Dataset and Task}
The StoryCommonsense dataset \cite{rashkin2018modeling} consists in short commonsense stories, with five natural language events each, and is annotated with the mental states of the characters - motivations and emotional reactions. To produce the dataset, annotators were asked to describe the mental states of the characters both using free-form natural language and emotion theory labels - Maslow's needs \cite{maslow1958dynamic}, Reiss' motives for motivations \cite{reiss2004multifaceted} and Plutchik basic emotions \cite{plutchik1980general}. The annotations describe how humans think about events and how they infer motives and emotions. People use a character-based approach to reason about their social worlds, which is reflected in this dataset. For that reason, we believe that the results obtained with this dataset transfer well to other interaction scenarios that we aim to model in the context of our project. 

The task in which our approach (see \sect \ref{subsec:expeval-approach}) is evaluated consists in labeling the emotional reactions of the characters in each story of the StoryCommonsense dataset. Particularly, we try to label each story event-character pair with a subset of the eight Plutchik basic emotions used to annotate the dataset - \textit{surprise}, \textit{disgust}, \textit{sadness}, \textit{joy}, \textit{anger}, \textit{fear}, \textit{trust}, \textit{anticipation}.

\subsection{Baselines}\label{sec:baselines}
%

The most straightforward approach to the task is to train a classifier that receives an unprocessed story event and a character and outputs its emotional reactions. Such classifiers were trained and tested by \citeauthor{rashkin2018modeling}\cite{rashkin2018modeling}, namely TF-IDF features, max-pooled GloVe embeddings, an LSTM and a CNN. The mentioned classifiers were trained both with and without a form of story context. Particularly, besides the event-character pair $(s_t, c_i)$, the classifiers are also inputted the previous events of the story where the character is explicitly mentioned. Event though results across all classifiers suggested the benefits of including this form of story context, the recursive input of story events is computationally expensive.  Although differently, we also include story context in our approach as described in, \sect~\ref{subsec:expeval-approach}. 

We also set as a baseline \citeauthor{bosselut2019dynamic} work \cite{bosselut2019dynamic}, who experimented COMET - a GPT language model previously fine-tuned on commonsense knowledge graph completion \cite{sap2019atomic} - on StoryCommonsense. The approach (\textit{COMET-CGA}) also included commonsense inference and was evaluated in a zero-shot setting, showing similar results to the  aforementioned supervised approaches. Additionally, in the few-shot setting, their approach outperformed the supervised baselines. Finally, the same authors used the GPT language model fine-tuned on StoryCommonsense, outperforming every baseline \cite{bosselut2019dynamic}. While producing better empirical results, the approach has increased computational costs and a bigger loss of generality when compared with the zero-shot setting. These limitations may constrain applications of the model to small datasets.

\subsection{Test Setting}

We test our approach under the zero-shot and few-shot setting and compare it with the previously discussed baselines. In order to show that our approach can be even further optimized to a specific task, beyond the one-shot setting, in a supervised manner, we also experiment fine-tuning the COMET model on a StoryCommonsense training set, consisting of $20\%$ of the development set, and skipping the commonsense inference step of the pipeline.

Table~\ref{tab:wordemolink} shows the dictionary we use to correspond the Plutchik emotions with words from COMET's vocabulary V. 

The metrics we use are Precision, Recall and F1-score. We give an example of how such metrics are computed by us in this task. Suppose that an event-character pair $(s_t, c_i)$ as an annotated set of emotions $Y_{s_t, c_i} = \{emotion_1, emotion_2, emotion _8\}$. Suppose additionally that, using CHARET, we predict that the set of emotions is instead $\hat{Y}_{s_t, c_i}=\{emotion_1, emotion_2, emotion_7\}$. In this case, we have 2 True Positives, 1 False Positive, 1 False Negative and 4 True Negatives. After summing each of these quantities across all event-character pairs, we compute the metrics we use.
\begin{table}[]
\caption{Plutchik emotion dictionary.}
\begin{tabular}{ll}
\hline
Emotion      & Word \\ \hline
Suprise      & surprised      \\
Disgust      & disgusted      \\
Sadness      & sad            \\
Joy          & happy          \\
Anger        & angry          \\
Fear         & fearful        \\
Trust        & trusting       \\
Anticipation & excited        \\ 
\hline
\end{tabular}
\label{tab:wordemolink}
\end{table}

\subsection{Results}\label{subsec:results}
We show the relative frequency at which each emotion is annotated for a character in a story event on the training set and show the results on Table~\ref{table:emotion_relative_freq}. These results are used to define the emotion-specific thresholds $k_y$ in the zero-shot setting.
\begin{table}[]
\caption{Relative frequency at which each emotion is annotated for a character in a story event on the training set.}
\begin{tabular}{llll}
\hline
Emotion & Actors & Objects \\ \hline
Surprise               & 38.8                     & 32.6                     \\
Disgust                & 18.3                     & 13.6                     \\
Sadness                & 25.2                     & 19.9                     \\
Joy                    & 53.0                     & 33.4                     \\
Anger                  & 19.3                     & 15.1                     \\
Fear                   & 26.3                     & 20.1                     \\
Trust                  & 34.1                     & 24.0                     \\
Anticipation           & 56.4                     & 33.7                     \\ 
\hline
\end{tabular}
\label{table:emotion_relative_freq}
\end{table}
%
%

Table~\ref{tab:results-main} shows the main results obtained on the task, compared with the discussed baselines. Our approach appears in the table with the name CHARET. The best results of each learning setting - zero-shot, few-shot and supervised- are bolden.
\begin{table}[]
\caption{Performance of ours and previous approaches to the StoryCommonsense emotion classification task.}
\begin{tabular}{llll}
\hline
\textbf{Model}              & Precision & Recall & F1    \\ \hline
\textbf{Zero-shot}          &           &        &       \\
Random                      & 20.6      & 20.8   & 20.7  \\ 
COMET - DynaGen             & \textbf{38.9}      & 39.3   & 39.1  \\
CHARET                      & 31.1      & \textbf{77.4}   & \textbf{44.3}  \\ 
\hline
\textbf{Few-shot} &         &        &       \\
COMET - DynaGen             & 31.2      & 65.1   & 42.2  \\
CHARET                      & \textbf{39.4}     & \textbf{81.5}  & \textbf{53.1} \\ \hline
\textbf{Supervised}         &           &        &       \\
TF-IDF                      & 20.1      & 24.1   & 21.9  \\
GloVe                       & 15.2      & 30.6   & 20.3  \\
LSTM                        & 20.3      & 30.4   & 24.3  \\
CNN                         & 21.2      & 23.4   & 22.2  \\
BERT                         & \textbf{65.6}      & 56.9   & \textbf{61.0}  \\ 
CHARET                       & 46.4      & \textbf{82.7}   & 59.5  \\ 
\hline
\end{tabular}
\label{tab:results-main}
\end{table}

The results indicate the benefits of our approach compared to others across all learning settings. Additionally, we notice that the performance increases as the level of task-specific knowledge increases - from zero-shot to few-shot; from few-shot to supervised.


\subsubsection{Errors introduced by the tools used for caracter role-labelling}

It is important to note that the pipeline proposed in this paper, may introduce some errors that may affect the overall performance of the approach. The reason is that each step of the pipeline uses tools that may not produce an accurate output at all times. The output of NeuralCoref feeds PredPatt, which in turn is responsible for creating the event-character pairs $(s_t, c_i)$. We evaluate how much error is introduced by the two tools used on labeling a character's role on a story event, NeuralCoref and PredPatt, on the training set. Table~\ref{table:tools_eval} shows the results.
\begin{table}[]
\caption{Performance of character role-labeling.}
\begin{tabular}{lll}
\hline
Precision & Recall & F1     \\ \hline
89.0    & 63.5 & 74.1 \\ \hline
\end{tabular}
\label{table:tools_eval}
\end{table}
The results show that our character-role labeling step is precise to some extent, as $89\%$ of the character we classify as actors of events are correctly classified. However, recall is not as high as we would like. This represents an additional venue for improvement. If we are able to improve the character-role labelling process (by using improved language models), we believe that we will be able to achieve better results in the downstream task of emotion tracking and prediction.

\section{Discussion}

 Our approach explored how data-driven and semantic tools combined allow to keep track of the emotional state of characters in a story [\textbf{RQ1}] the way humans do. The results indicate that our character-based approach to emotion inference outperforms previous machine learning approaches in a dataset annotated for concepts that people use to decode social interactions. 

By revisiting the sub-research questions in \sect \ref{sec:research_q}, we can positively say that by considering the semantics of emotions and thus, conducting  semantic-aware reasoning, we can make a better use of a commonsense inference tool that has links between causes and effects. Emotion classification is a task that entails many other pre-processing steps (\eg personality recognition, anaphora resolution, polarity detection, \etc) that are not considered in this work. Yet, our work points out that an approach that combines data and semantic analysis should be considered in order to create more believable interactions between IVAs and humans. At the same time, in the real-world applications, we should expect to have more noisy data points, with a poorer structure, which could result in poorer results. Predicate and argument extraction, the backbone of our approach, may be more challenging under those conditions as shown in open-information scenarios.

\section{Conclusion}

Advances in NLP have propelled the use of Intelligent Virtual Agents in a myriad of contexts, in particular in the role of automated assistants. These systems are a good example that data-driven approaches to human-agent interaction alone, are not sufficient to produce socially resonant behavior.

We proposed CHARET, a character centered approach approach that leverages on a simple form of semantic role labelling and state-of-the-art commonsense and processing tools for the task of emotion tracking and prediction from events. We validated our approach on a well structured dataset with clear cut events that represent a coherent story. Our approach outperfoms previous works that do not account for the semantics and subjectiveness of emotions when inferring the emotional state of characters. Although not yet ideal, the results obtained for the few-shot setting - requiring a small amount of training - reached an F1 score of 53.1, a 25.8\% increase over the baseline COMET - DynaGen. These results are promising, as they suggest that a layered approach to emotion tracking and prediction can yield better results than end-to-end approaches.  

In future work we intend to explore whether this approach performs well in more challenging domains, namely Fairy Tales. Additionally, we consider including other aspects that define a character and a situation to further capture emotional content from events. We are also interested in exploring if the same commonsense tools facilitate this type of inferences and what are their limitations.




\begin{acks}
This work was partially supported by national funds through Fundação para a Ciência e Tecnologia under project SLICE with reference PTDC/CCI-COM/30787/2017, University of Lisbon and Instituto Superior Técnico and INESC-ID multi annual funding with reference UIDB/50021/2020. 
\end{acks}



\bibliographystyle{ACM-Reference-Format} 


\end{document}